\documentclass[runningheads]{llncs}
\usepackage{graphicx}
\usepackage{amsmath}
\usepackage{newtxmath}
\usepackage{amsfonts}
\usepackage{multirow}
\usepackage[dvipsnames]{xcolor}
\usepackage{amsmath}
\usepackage[pagebackref=true,breaklinks=true,colorlinks,bookmarks=false,letterpaper=true]{hyperref}

\usepackage{booktabs}

\usepackage{color}

\begin{document}
\title{Task-Oriented Low-Dose CT Image Denoising}

\titlerunning{MICCAI'21 Paper ID: 1982}

\authorrunning{MICCAI'21 Paper ID: 1982}

%

%
\titlerunning{Task-Oriented Low-Dose CT Image Denoising}

\author{Jiajin Zhang $^\star$
\and Hanqing Chao
\thanks{J.Z. and H.C. are co-first authors.}
\and Xuanang Xu
\and Chuang Niu
\and Ge Wang
\and Pingkun Yan
\thanks{Corresponding author.}}


\authorrunning{J. Zhang et al.}
%
\institute{Department of Biomedical Engineering and Center for Biotechnology and Interdisciplinary Studies, Rensselaer Polytechnic Institute, Troy, NY 12180, USA\\
\email{\{zhangj41, chaoh, xux12, niuc, wangg6, yanp2\}@rpi.edu} 
}
\maketitle              
\begin{abstract}

The extensive use of medical CT has raised a public concern over the radiation dose to the patient. Reducing the radiation dose leads to increased CT image noise and artifacts, which can adversely affect not only the radiologists judgement but also the performance of downstream medical image analysis tasks.
Various low-dose CT denoising methods, especially the recent deep learning based approaches, have produced impressive results. However, the existing denoising methods are all downstream-task-agnostic and neglect the diverse needs of the downstream applications. 
In this paper, we introduce a novel Task-Oriented Denoising Network (TOD-Net) with a task-oriented loss leveraging knowledge from the downstream tasks. Comprehensive empirical analysis shows that the task-oriented loss complements other task-agnostic losses by steering the denoiser to enhance the image quality in the task related regions of interest. Such enhancement in turn brings general boosts on the performance of various methods for the downstream task. The presented work may shed light on the future development of context-aware image denoising methods. 
Code is available at \url{https://github.com/DIAL-RPI/Task-Oriented-CT-Denoising_TOD-Net}.

\keywords{Low-dose CT \and Image denoising \and Task-oriented loss \and downstream task.}
\end{abstract}
\section{Introduction}

Computed tomography (CT) is one of the most important breakthroughs in modern medicine, which is routinely used in hospitals with millions of scans performed per year world wide. However, the ionizing radiation associated with CT puts patients at the risk of genetic damage and cancer induction. Low-dose CT (LDCT) uses lower radiation dose for imaging, which helps reduce the risks but increases noise and artifacts in reconstructed images.
These compromised CT images with artifacts can affect not only the diagnosis by physicians but also the downstream medical image processing tasks. 

In order to tackle this problem, the research community has developed various LDCT denoising methods. They can be grouped into the three categories, \textit{i.e.}, sinogram filtration~\cite{manduca2009projection,wang2012image,wang2005sinogram}, iterative reconstruction~\cite{beister2012iterative,hara2009iterative}, and image post-processing~\cite{chen2013improving,feruglio2010block,yang2018low}. 
Classic post-processing methods include K-SVD~\cite{chen2013improving}, non-local means (NLM)~\cite{ma2011low} and block-matching 3D (BM3D)~\cite{kang2013image,feruglio2010block}.
With the renaissance of artificial intelligence in the past decade, various deep neural networks were proposed to denoise LDCT images, which became the main stream methods. 
Chen et al.~\cite{chen2017low} first presented a 3-layer convolution neural network (CNN) trained to minimize the mean square error (MSE) for LDCT post denoising. Yang et al.~\cite{yang2018low} introduced a Wasserstein Generative Adversarial Network (WGAN) with a perceptual loss to keep more detailed information in denoising. Most recently, the self-attention mechanisms and self-supervised learning methods have been introduced in the field to further improve the performance~\cite{li2020sacnn}. Aiming at alleviating the difficulties brought by the noise to downstream tasks, this work also focuses on deep learning based post-processing methods for both image denoising and downstream task performance improvement.

Although many efforts have been made in LDCT denoising, existing methods are all downstream-task-agnostic. Specifically, deep learning based denoising methods all intend to reduce a distance between denoised LDCT images and the corresponding normal-dose CT (NDCT) counterparts. The diverse needs from the downstream tasks have been largely overlooked.
In this paper, we argue that the denoising module should be trained with the awareness of downstream applications. This will bring two major benefits. One is that using the downstream task requirements can enhance the denoising performance on those task-related regions. The other is that this image quality enhancement can in turn boost the performance of the downstream tasks.
To achieve this goal, here we propose a novel Task-Oriented Denoising Network (TOD-Net) with a task-oriented loss in the WGAN framework. Specifically, in training the TOD-Net, we incorporate a fixed representative network of the downstream task and use its task specific loss to guide the optimization of the TOD-Net's parameters. We demonstrate that in the whole workflow, the TOD-Net can not only significantly improve the performance of the specific network in the training phase but also generally boost the performance of various other networks for the same downstream task. Experiments on two datasets show that the image quality of the TOD-Net on the task-related regions is clearly superior to that of the other denoising methods.

The contributions of this paper can be summarized in two aspects:
1) This is the first work leveraging the knowledge of downstream tasks in LDCT denoising. Compared with existing task-agnostic denoising methods, the proposed Task-Oriented Denoising Network (TOD-Net) can significantly improve the denoised CT image quality on the task-related regions.
2) With the targeted image quality improvement, integrating TOD-Net with a downstream model can lead to a significant performance boost on the downstream task. In addition, TOD-Net can significantly improve the performance of various other networks for the same downstream task.

\section{Method}

\begin{figure*}[t]
\centering
\includegraphics[width=\textwidth, clip=true, trim=30 50 25 35]{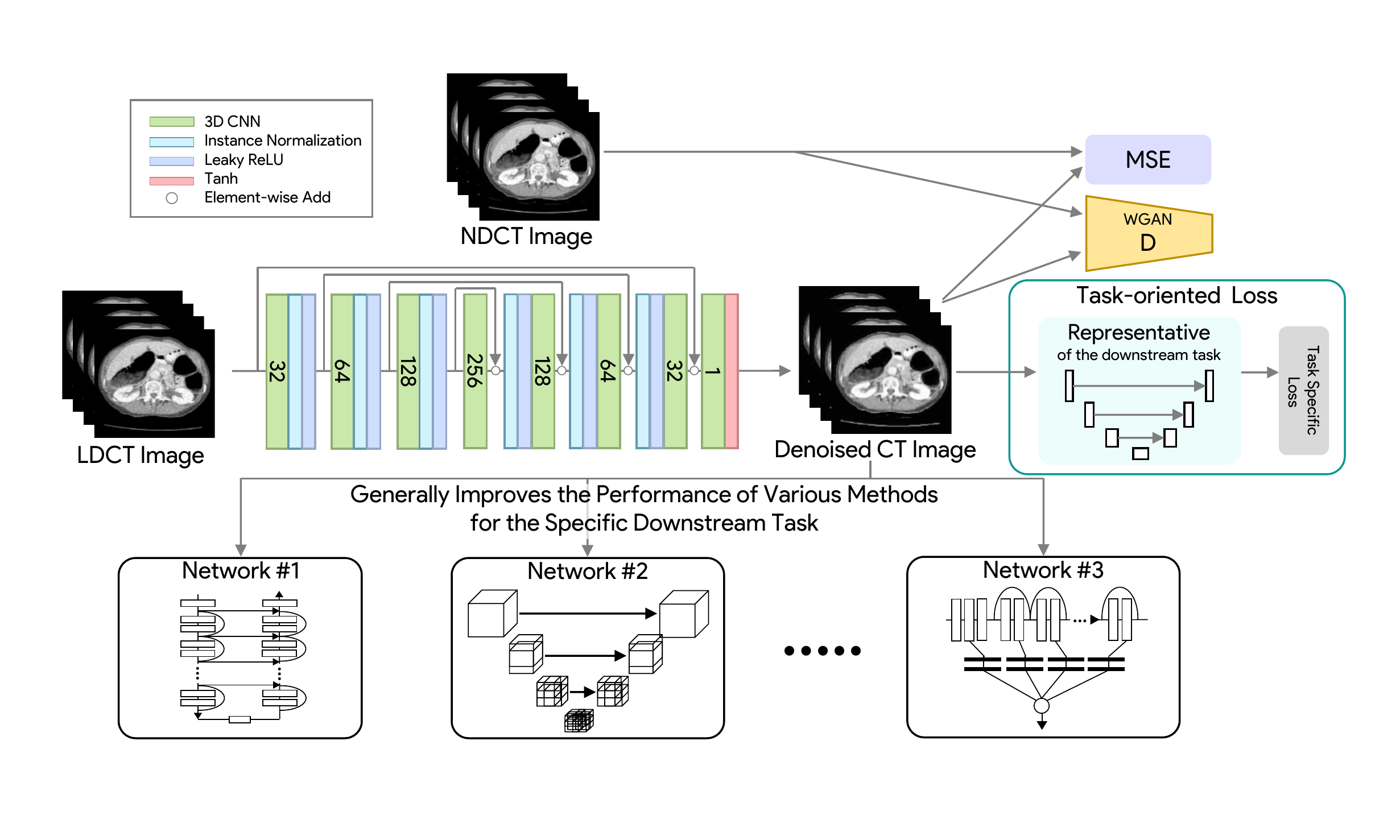}
\caption{\textbf{Illustration of the proposed TOD-Net.} The proposed TOD-Net consists of a denoiser and 3 loss estimators. The size of the 3D kernels in the denoiser is $3\times3\times3$. The numbers in the 3D CNN block are the channel number. 
}
\label{fig:TOD-Net}
\end{figure*}

The proposed TOD-Net generates denoised image from LDCT guided by a unique combination of three losses. These losses include an mean square error (MSE) loss to control the global Euclidean distance between a generated image and its NDCT counterpart, a WGAN discriminator to shrink the distance bewteen their distributions, and a Task-oriented loss to reflect downstream task-related requirements. Fig.~\ref{fig:TOD-Net} shows an overview of our TOD-Net.

\subsection{WGAN for LDCT Denoising} 

The TOD-Net is based on the well-known WGAN~\cite{arjovsky2017wasserstein}. Let $\mathbf{x}$ denote a LDCT image and $\mathbf{x}^*$ be its counterpart NDCT image. WGAN consists of a denoiser $G$ and a discriminator $D$ to generate an image $\hat{\mathbf{x}}=G(\mathbf{x})$ with a distribution of $\mathbf{x}^*$. 

Denoting the distribution of $\mathbf{x}$ as $Q$ and the distribution of $\mathbf{x}^*$ as $P$, the optimization of loss functions of the WGAN can be formulated as:
\begin{equation}
\label{eq:gan_d}
L_D(\theta_D) = \mathbb{E}_{\mathbf{x}^*\sim{P}}[D(\mathbf{x}^*;\theta_D)]-\mathbb{E}_{\mathbf{x}\sim{Q}}[D(G(\mathbf{x};\theta_G);\theta_D)];
 \text{subject\ to}\ ||\theta_D||_{1} \le \epsilon,
\end{equation}
\begin{equation}
\label{eq:wgan2}
L_{GAN}(\theta_G) = \mathbb{E}_{\mathbf{x}\sim{Q}}[D(G(\mathbf{x};\theta_G);\theta_D)],
\end{equation}
where $\theta_D$ and  $\theta_G$ are the parameters of $D$ and $G$ respectively. Instead of using the Jensen–Shannon (JS) divergence or Kullback-Leibler (KL) divergence to discriminate $P$ and $Q$, the discriminator of WGAN applies a more smooth measurement, Wasserstein distance, to evaluate the discrepancy between $P$ and $Q$ which can effectively stabilize the training of the network.

\subsection{Analysis of Task-oriented Loss}
\label{sec:tdloss}
The proposed task-oriented loss, our main innovation, is to equip the denoiser with an explicit awareness of the downstream task in the training phase. Under this task-oriented guidance, the denoising network will enhance the specific features essential to the downstream task and boost the whole medical image processing pipeline's performance. Specifically, a representative model $T$ of a downstream task is incorporated into the training process. As shown in Fig.~\ref{fig:TOD-Net}, the denoised images are fed to $T$ to compute the task-oriented loss. In this paper, we demonstrate our TOD-Net based on the medical image segmentation task. Therefore, we minimize the prevalent Dice loss to compute the task-oriented loss
\begin{equation}
\label{eq:tdloss}
L_{t}(\theta_G)=\mathbb{E}_{\mathbf{x}\sim{Q}}[1-Dice(T(G(\mathbf{x};\theta_G)))].
\end{equation}

\setlength{\textfloatsep}{1 \baselineskip plus 0\baselineskip minus 0.5\baselineskip}
\begin{figure*}[t]
\centering
\includegraphics[width=\textwidth, clip=true, trim=50 80 50 70]{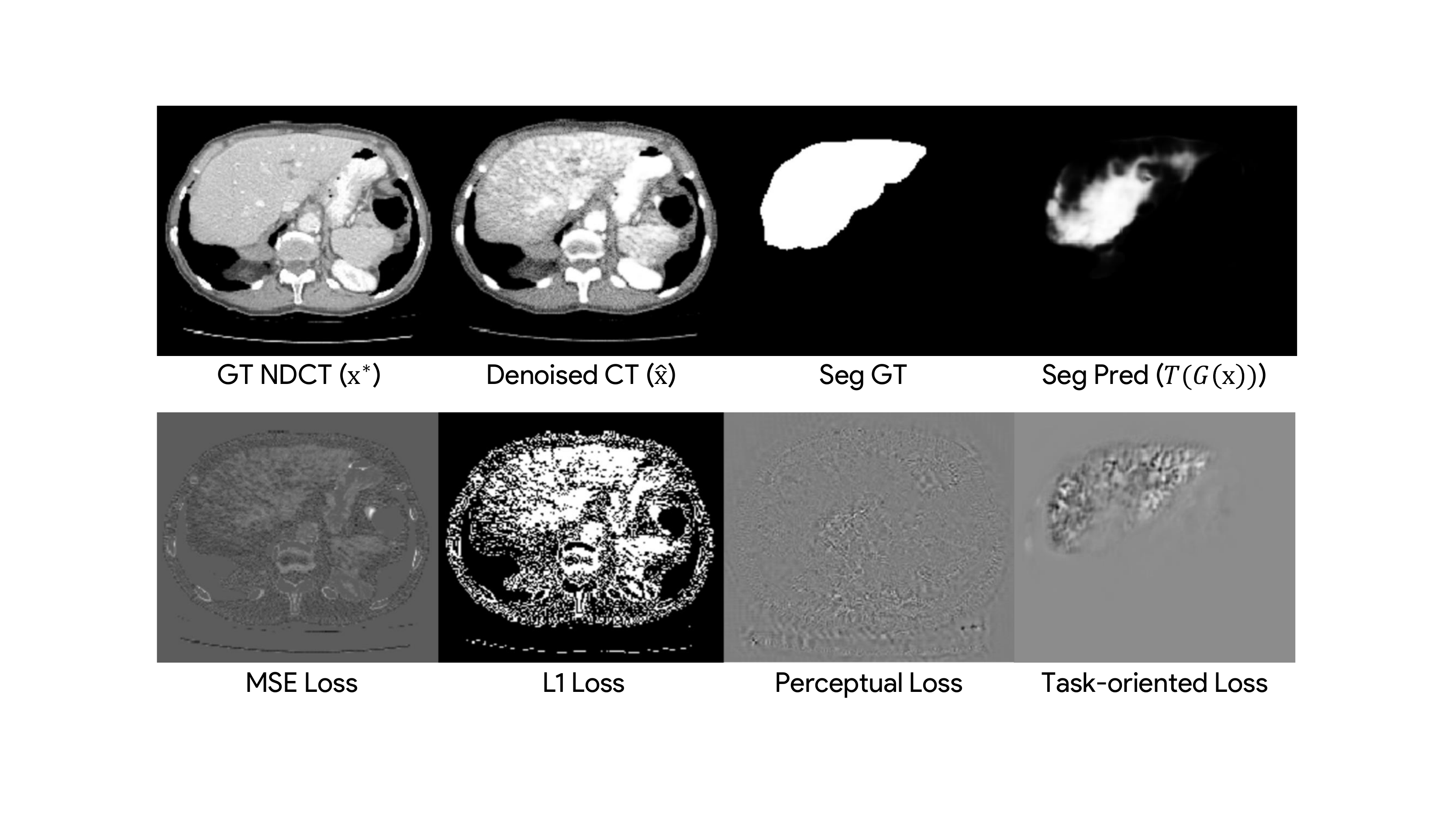}
\caption{\textbf{Gradient maps of different losses.} The first row shows ground truth NDCT image $\mathbf{x}^*$, the output of the denoising network $\hat{\mathbf{x}}$, the segmentation ground truth, and the out put of the representative model incorporated in the task-oriented loss. The second row shows the map of $\frac{\partial L}{\partial \hat{\mathbf{x}}}$ of 4 different losses.}
\label{fig:grad}
\end{figure*}

To analyze how the task-oriented loss complements other commonly used denoising losses, we examine the similarities and differences between their partial derivatives with respect to the denoiser's parameters $\theta_G$. Specifically, we compare the task-oriented loss to MSE loss $L_{MSE}$, $L_1$ loss, and perceptual loss ($L_p$). For brevity, we use $\hat{\mathbf{x}}$ to represent $G(\mathbf{x})$. Then we have
\begin{align}
\label{eq:g_td}
\frac{\partial L_t}{\partial \theta_G} = \frac{\partial L_t}{\partial \hat{\mathbf{x}}}\frac{\partial \hat{\mathbf{x}}}{\partial \theta_G},& \\
L_{MSE} = \frac{1}{2}||\hat{\mathbf{x}}-\mathbf{x}^*||_2^2;  \frac{\partial L_{MSE}}{\partial \theta_G} = ||\hat{\mathbf{x}}-\mathbf{x}^*||_2\frac{\partial \hat{\mathbf{x}}}{\partial \theta_G},& \\
\label{eq:g_l1}
L_1 = ||\hat{\mathbf{x}}-\mathbf{x}^*||_1; 
\frac{\partial L_1}{\partial \theta_G} = \vmathbb{1}[\hat{\mathbf{x}}-\mathbf{x}^*]\frac{\partial \hat{\mathbf{x}}}{\partial \theta_G},& \\
\label{eq:g_lp}
L_p = \frac{1}{2}||f(\hat{\mathbf{x}})-f(\mathbf{x}^*)||^2_2;  \frac{\partial L_p}{\partial \theta_G} = ||f(\hat{\mathbf{x}})-f(\mathbf{x}^*)||_2\frac{\partial f}{\partial \hat{\mathbf{x}}}\frac{\partial \hat{\mathbf{x}}}{\partial \theta_G},&
\end{align}
where $f(\cdot)$ denotes the network used in the perceptual loss. In Eq.~\ref{eq:g_l1}, $\vmathbb{1}[\cdot]=1$ when $\cdot \ge 0$ and  $\vmathbb{1}[\cdot]=-1$ when $\cdot < 0$. From Eqs.~\ref{eq:g_td}-\ref{eq:g_lp}, we can see that the partial derivatives of all these four losses share the same term of $\partial \hat{\mathbf{x}} / {\partial \theta_G}$. The differences between these loss functions lie only in the other part of the derivatives, denoted by ${\partial L}/{\partial \hat{\mathbf{x}}}$. For intuitive understanding, Fig.~\ref{fig:grad} visualizes the values of ${\partial L}/{\partial \hat{\mathbf{x}}}$ calculated on an output $\hat{\mathbf{x}}$ generated by a half-trained denoiser, \textit{i.e.}, TOD-Net before converging. Liver segmentation is the downstream task in this demonstration.
Since the decoder has not been fully trained,
the ``denoised'' image $\hat{\mathbf{x}}$ includes many artifacts, which significantly degrade the performance of the downstream segmentation network. The second row shows the differences between the task-driven loss and the other three losses, with the partial derivatives of the task-oriented loss being focused on the liver. Such an attention to the task-related regions steers the denoiser to improve the image quality in these regions. Since the improvements are not for a specific representative network the improvements can enhance various other methods for this task.

\subsection{Training Strategy}
\label{sec:TS}
The total loss used for optimizing the denoiser $G$ is given by
\begin{equation}
\label{eq:Gloss}
L_{G} = L_{GAN} + L_{t} + \lambda L_{MSE},
\end{equation}
where $\lambda$ is a hyper-parameter to weigh the MSE loss. The task-oriented loss and the MSE loss work together to enhance the local features required by the downstream task, while maintaining a high global quality. The loss for training the discriminator $D$ is denoted as $L_D$ as defined in Eq.~\ref{eq:gan_d}. In each training iteration, the two losses $L_D$ and $L_G$ are updated alternatively.

\section{Experiments}
The proposed TOD-Net was evaluated on two datasets from two aspects: 1) image quality of the denoised image, and 2) performance of the downstream task taking the denoised image as the input. The image quality is measured by root-mean-square error (RMSE), structural similarity index (SSIM), and peak signal-to-noise ratio (PSNR). The performance of the downstream task, medical image segmentation, is evaluated by the Dice score. Our model was compared with three baselines, VGG-WGAN~\cite{yang2018low} SA-WGAN~\cite{li2020sacnn} and MSE-WGAN. We trained the VGG-WGAN and SA-WGAN on the two datasets with the exact procedure and hyper-parameters used in their original papers. The MSE-WGAN has the same structure as the TOD-Net but without the task-oriented loss. It was trained with the same protocol as the TOD-Net.

\subsection{Datasets}

We evaluated our model on two publicly available datasets for single-organ segmentation, \textit{i.e.}, LiTS (Liver tumor segmentation challenge)~\cite{bilic2019liver}, and KiTS (Kidney tumor segmentation challenge)~\cite{heller2019kits19}.
LiTS consists of 131 abdominal NDCT images collected from different hospitals and the in-plane resolution varies from 0.45mm to 6mm and the slice spacing is between 0.6mm and 1.0mm. 
We split the dataset into training set, validation set and test set with a ratio of 70\% (91 images), 10\% (13 images), and  20\% (27 images). 
KiTS includes abdominal NDCT images from 300 patient, 210 of which are publicly available.
Since the number of samples is relatively large, we split them into training, validation and test set with a ratio of 60\% (126 images), 20\% (42 images), and  20\% (42 images).

To train and test the proposed TOD-Net, we synthesized the low-dose counterparts of all NDCT images in the two datasets using an industrial CT simulator, CatSim by GE Global Research~\cite{deman2007catsim}, which incorporates finite focal spot size, realistic x-ray spectrum, finite detector pitch and other data imperfections, scatter with and without an antiscatter grid, and so on. We denote these two LDCT datasets as LD-LiTS and LD-KiTS in the following sections. 

\subsection{Segmentation Networks}
In this paper, four different segmentation networks were used as our downstream models, including U-Net \cite{ronneberger2015u}, V-Net\cite{milletari2016v}, Res-U-Net \cite{diakogiannis2020resunet} and Vox-ResNet \cite{chen2018voxresnet}. Each segmentation network was pretrained independently on the LiTS and KiTS training sets.
On the LiTS test set, the four segmentation networks in the above order achieved Dice of 94.31\%, 92.80\%, 93.51\%, and 92.08\%, respectively. On the KiTS test set, these four segmentation networks achieved Dice of 90.87\%, 89.61\%, 89.78\%, and 89.35\%, respectively. 
All the four networks did NOT achieved a SoTA Dice score ($> 95\%$), because our data split is different from the original challenge. We split the challenge training set into training, validation, and test sets. The segmentation models are trained with fewer data and evaluated on a different test set than the original challenge. 


\subsection{Implementation Details}
For all the datasets, the pixel intensity is clamped with a window of [-200,200]HU and normalized to [0,1]. In Sec. 3.1, to train a denoising network, we first resized the images to have an in-plane resolution of 1.5mm$\times$1.5mm and a slice spacing of 1mm. Then, the axial slices are center cropped/zero padded to a size of 256$\times$256 pixels. To generate training batches, the 3D LDCT images and segmentation ground truth are split into multiple overlapping 256$\times$256$\times$32 sub-volumes. In the validation and test phase, the TOD-Net works directly on the original volume with size of 512$\times$512$\times$\#slices.

When training our TOD-Net, only the above trained U-Net \cite{ronneberger2015u} was used as the representative downstream model in the task-oriented loss. Then, the trained TOD-Net was directly applied to all the four segmentation networks to verify the generalizability. 
%
%
The WGAN discriminator used in the training is a 3 layers 3D CNN followed by a fully connected layer with a single output. Each 3D convolutional layer is followed by a batch normalization layer and a leaky ReLU layer. The kernel size of all convolutional layers is 3$\times$3$\times$3. The channel sizes of the layers are 32, 64, and 128, respectively.
%
%
Due to the limitation of GPU memory, in the training phase of TOD-Net, we further cropped 3D LDCT image into 256$\times$256$\times$32 patches with a sliding window. On LD-LiTS and LD-KiTS, the TOD-Net was separately trained by the RMSprop optimizer with a learning rate of 0.0005 and a batch size of 4 for 50 epochs. The hyperparameter $\lambda$ in $L_G$ was set to be 0.5 and the gradient clipping value $\epsilon$ in Eq.~\ref{eq:gan_d} was set to be 0.01. One checkpoint was saved at the end of each epoch and the checkpoint with the best performance on the validation set was used as the final model.





\subsection{Enhancement on task-related Regions}
\label{sec:img_quality}
%
We first quantitatively evaluated the denoising quality of the proposed TOD-Net. We computed SSIM, RMSE and PSNR on both whole image and regions of interest (ROIs) of the downstream task, 
i.e., liver for LiTS and kidney for KiTS.
Results are shown in Table~\ref{tab:quality}. Compared with the other two WGAN denoisers, TOD-Net ranked the top in all evaluation metrics at both whole image and ROI levels.

\begin{table}[t]
	\centering
	\caption{Image quality analysis of denoised LDCT images on LiTS and KiTS datasets.}
	\label{tab:quality}
	\scalebox{0.86}{
	\begin{tabular}{l||c|c|c||c|c|c||c|c|c||c|c|c}
	    \toprule
	    \multirow{3}{*}{\textbf{Denoiser}} & \multicolumn{6}{c||}{\textbf{LD-LiTS}}& \multicolumn{6}{c}{\textbf{LD-KiTS}}
		\\ \cline{2-13}
		& \multicolumn{3}{c||}{\textbf{ROIs}} &  
		\multicolumn{3}{c||}{\textbf{Whole Image}} & \multicolumn{3}{c||}{\textbf{ROIs}} & \multicolumn{3}{c}{\textbf{Whole Image}}\\
		\cline{2-13}
		& SSIM & RMSE & PSNR & SSIM & RMSE & PSNR & SSIM & RMSE & PSNR & SSIM & RMSE & PSNR\\
		\midrule
		VGG-WGAN & 29.5 & 53.0 & 17.7 & 72.3 & 32.9 & 22.3 & 46.7 & 58.3 & 16.9 & 67.4 & 37.8 & 21.1\\
		MSE-WGAN & 35.1 & 41.8 & 19.8 & 74.7 & 31.8 & 22.3 & 50.0 & 53.5 & 17.6 & 65.9 & 37.0 & 21.9 \\
		SA-WGAN & 37.5 & 40.1 & 20.1 & 73.1 & 31.1 & 22.7 & 54.5 & 53.7 & 18.3 & 67.2 & 34.9 & \textbf{22.0} \\
		TOD-Net & \textbf{40.7} & \textbf{35.4} & \textbf{21.3} & \textbf{76.7} & \textbf{28.5} & \textbf{23.3} & \textbf{61.1} & \textbf{45.5} & \textbf{19.0} & \textbf{69.3} & \textbf{32.9} & \textbf{22.0} \\
		\bottomrule
	\end{tabular}
	}
\end{table}

It is worth noting that image quality enhancement of TOD-Net on ROIs is even more significant than on whole images. This verifies that with information from the pretrained downstream model, the task-oriented loss locally promotes image quality of task-related regions.
Combining it with the MSE loss for global regularization, TOD-Net achieved the best performance both locally and globally.
%

\subsection{Boosting Downstream Task Performance}
\label{sec:boost}
\begin{table}[t]
	\centering
	\caption{Comparison of TOD-Net and other denoising models on the downstream task Dice score (\%) and generalizability on \textbf{(top)} LD-LiTS and \textbf{(bottom)} LD-KiTS.}
	\label{tab:dice_lits}
	\scalebox{0.96}{
	\begin{tabular}{l|c||c|c|c}
		\toprule
		\multirow{2}{*}{\textbf{Denoiser}} & \multirow{2}{*}{\textbf{U-Net}} & \multicolumn{3}{c}{\textbf{Generalize to Other Seg. Models}}\\
		\cline{3-5}
		& & \textbf{Vox-ResNet} & \textbf{V-Net} & \textbf{Res-U-Net} \\
		\midrule \midrule
		No denoiser & 88.75 & 79.82 & 89.75 & 90.36 \\
		VGG-WGAN & 92.34 $(p=0.006)$ & 90.44 $(p=0.002)$  & 91.21 $(p=0.004)$ & 91.82 $(p=0.004)$ \\
		MSE-WGAN & 92.45 $(p=0.005)$ & 89.49 $(p<0.001)$
        & 91.76 $(p=0.005)$ & 91.95 $(p=0.005)$ \\
		SA-WGAN & 92.99 $(p=0.007)$ & 89.26 $(p<0.001)$
        & 91.37 $(p=0.007)$ & 91.46 $(p=0.005)$ \\
		TOD-Net & \textbf{93.91} & \textbf{91.86} & \textbf{92.44} & \textbf{92.77} \\ \midrule
		\midrule
		No denoiser & 75.30 & 43.09 & 80.75 & 80.57 \\
		VGG-WGAN & 89.43 $(p=0.044)$ & 86.88 $(p<1e-3)$
        & 88.06 $(p=0.005)$ &  89.67 $(p=0.004)$\\
		MSE-WGAN & 88.82 $(p=0.004)$ & 86.58 $(p<0.001)$ & 88.46 $(p=0.004)$ & 88.66 $(p=0.004)$ \\
		SA-WGAN & 88.87 $(p=0.005)$ & 86.39 $(p<0.001)$
        & 88.66 $(p=0.006)$ & 88.75 $(p=0.005)$ \\
		TOD-Net & \textbf{90.21} & \textbf{89.81} & \textbf{89.83} & \textbf{90.03} \\
		\bottomrule
	\end{tabular}
	}
\end{table}

In this experiment, we cascaded the TOD-Net with each of the four segmentation networks to evaluate its influence on the downstream task. The results on LD-LiTS and LD-KiTS are shown in Table~\ref{tab:dice_lits}. 
Since LDCT images contains more noise and artifacts than NDCT images, directly applying segmentation models trained on NDCT to LDCT images would cause a significant performance degradation, as shown by the rows of ``No denoiser'' in Table~\ref{tab:dice_lits}.

In contrast, using the denoised images improved the segmentation performance. 
%
Measured by Dice scores, TOD-Net significantly outperformed all other denoising models ($p<0.05$) not only on the U-Net, which was used in the training of TOD-Net, but also on all three other segmentation networks. Such a generalizability can be explained by the significant image quality enhancement on the task-related regions mentioned earlier in Sec.~\ref{sec:img_quality}.

Since real clinical datasets including paired LDCT and NDCT with segmentation annotations are hard to obtain, we evaluate the performance of the TOD-Net on real CT images by applying TOD-Net on the original LiTS and KiTS datasets. The results in Table \ref{tab:real_ct} show that TOD-Net brings significant improvement for all segmentation networks on KiTS. On LiTS, all the downstream networks perform better except the performance of Res-U-Net decrease a little with non-significant difference ($p>0.05$).

\begin{table}[t]
	\centering
	\caption{TOD-Net performance on real CT images on the downstream task Dice score(\%) and generalizability on \textbf{(top)} LiTS and \textbf{(bottom)} KiTS.}
	\label{tab:real_ct}
	\scalebox{0.96}{
	\begin{tabular}{l|c||c|c|c}
		\toprule
		\multirow{2}{*}{\textbf{Denoiser}} & \multirow{2}{*}{\textbf{U-Net}} & \multicolumn{3}{c}{\textbf{Generalize to Other Seg. Models}}\\
		\cline{3-5}
		& & \textbf{Vox-ResNet} & \textbf{V-Net} & \textbf{Res-U-Net} \\
		\midrule \midrule
		No denoiser & 94.33 & 92.19 & 92.82 & \textbf{93.51} \\
		TOD-Net & \textbf{95.01}$(p=0.04)$ & \textbf{92.43}$(p=0.06)$ & \textbf{93.44}$(p=0.04)$ & 93.21$(p=0.08)$  \\ \midrule
		\midrule
		No denoiser & 90.93 & 89.39 & 89.58 & 89.81 \\
		TOD-Net & \textbf{91.52}$(p=0.03)$ & \textbf{90.41}$(p=0.01)$ & \textbf{ 91.37}$(p=0.01)$ & \textbf{91.03}$(p=0.03)$ \\
		\bottomrule
	\end{tabular}
	}
\end{table}

\begin{figure*}[t]
\centering
\includegraphics[width=\textwidth, clip=true, trim=0 0 0 0]{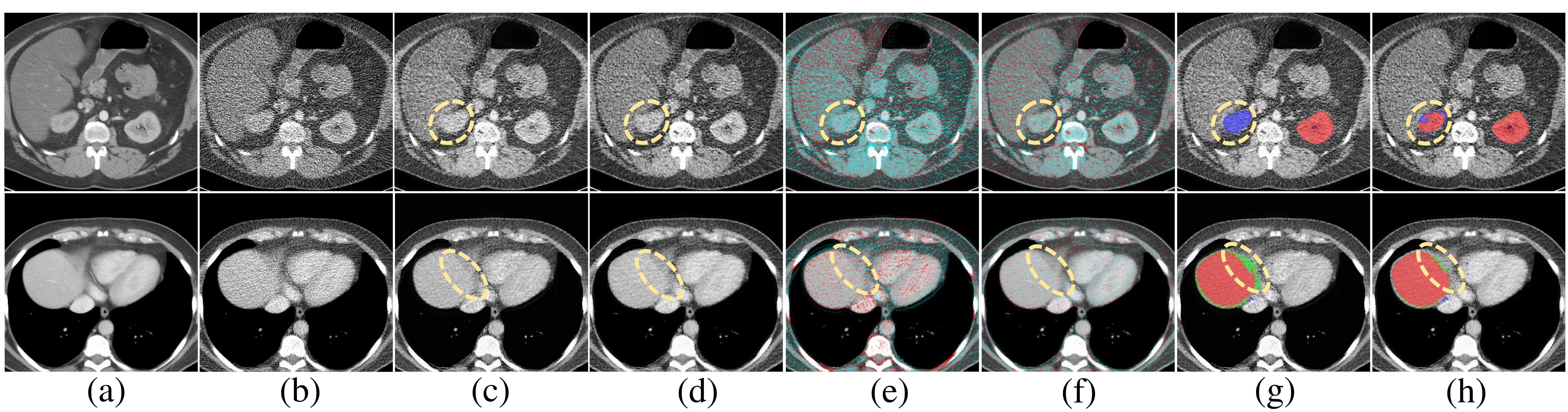}
\caption{\textbf{Case study.} 
 Slices selected from \textbf{(top row)} KiTS and \textbf{(bottom row)} LiTS. From left to right, each column shows \textbf{(a)} NDCT, \textbf{(b)} LDCT, denoised image with \textbf{(c)} MSE-WGAN,\textbf{(d)} TOD-Net, \textbf{(e)} difference map between (a) and (c), \textbf{(f)} difference map between (a) and (d)(Red and green represent error above zero and below zero, respectively), \textbf{(g)} segmentation results of (c) with U-Net, \textbf{(h)} segmentation results of (d) with U-Net. The red, blue and green depict the true positive, false negative, and false positive regions, respectively.
}
\label{fig:visual}
\end{figure*}

In addition to the quantitative analysis to demonstrate the high performance and great generalizability of TOD-Net, we further visualized the differences brought by the task-oriented loss.
%
Fig.~\ref{fig:visual} shows two example cases. It can be seen that, in columns (e) and (f), the errors of TOD-Net are significantly less than MSE-WGAN, especially on the ROIs, which leads to better segmentation results in Fig.~\ref{fig:visual}(h) compared to Fig.~\ref{fig:visual}(g).

\section{Conclusion}
In conclusion, we proposed a TOD-Net with a task-oriented loss for LDCT image denoising. Quantitative evaluations and case studies show that introducing the knowledge of downstream applications into the training of the denoising model can significantly enhance the denoised image quality especially in the downstream task-related regions. Such a targeted improvement of image quality in turn boosts the performance of not only the downstream model used for training the denoising model but also various other independent models for the same task. The presented work may potentially enable a new group of image denoising methods as well as performance enhancement of existing tasks.
In our future work, we will investigate the performance our task-oriented loss on other downstream tasks in medical image analysis.
%
%
%


\bibliographystyle{splncs04}
\bibliography{ref}

\end{document}